\documentclass[twocolumn,showpacs,preprintnumbers,amsmath,amssymb]{revtex4}
\usepackage{graphicx}
\usepackage{dcolumn}
\usepackage{bm}

\begin{document}

\preprint{cond-mat/0405298}

\title{First-principles prediction of a decagonal quasicrystal
containing Boron}

\author{M. Mihalkovi\v{c}}
\altaffiliation[Also at ]{Institute of Physics, Slovak Academy of Sciences, 84228 Bratislava, Slovakia}
\author{M. Widom}
\affiliation{
Department of Physics, Carnegie Mellon University\\
Pittsburgh, PA  15213
}

\date{\today}

\begin{abstract}
We interpret experimentally known B-Mg-Ru crystals as quasicrystal
approximants.  These approximant structures imply a deterministic
decoration of tiles by atoms that can be extended quasiperiodically.
Experimentally observed structural disorder corresponds to phason
(tile flip) fluctuations.  First-principles total energy calculations
reveal that many distinct tilings lie close to stability at low
temperatures.  Transfer matrix calculations based on these energies
suggest a phase transition from a crystalline state at low
temperatures to a high temperature state characterized by tile
fluctuations.  We predict B$_{38}$Mg$_{17}$Ru$_{45}$ forms a decagonal
quasicrystal that is metastable at low temperatures and may be
thermodynamically stable at high temperatures.
\end{abstract}

\maketitle

While evaluating the stability of crystalline phases competing with
metallic glass formation~\cite{bfezry_unp}, we discovered a number of
previously unrecognized decagonal quasicrystal
approximants~\cite{mrs03}.  These are ordinary, though complex,
crystals whose local structural motifs may be naturally extended to
form a quasiperiodic structure with an axis of 10-fold rotational
symmetry.  We created a series of hypothetical quasicrystal
approximant structures based on these motifs and calculated their
total energy.  Based on our findings we propose that
B$_{38}$Mg$_{17}$Ru$_{45}$ should posses a decagonal quasicrystal
state that is at least metastable, and potentially even
thermodynamically stable at high temperature.

This prediction is noteworthy because: (1) the compound contains a
substantial amount of Boron, which has been speculated to form
quasicrystalline structures~\cite{Henley_B,Kimura,Boustani,Weygand}
though none are yet known; (2) the predicted structure is quite
different from the established structures of Al-rich decagonal
quasicrystals~\cite{other_deca,MM8} in that it is an {\em intrinsic}
ternary, while known Al-rich decagonals are essentially {\em
pseudobinaries}; (3) we predict both structure and {\em existence}
from first-principles, while in prior work~\cite{MM8}, existence was
taken as a crucial experimental input for structure prediction; (4)
first-principles total energy calculations confirm the feasibility of
entropic stabilization~\cite{entropic}.

\begin{figure}
\includegraphics[width=3in]{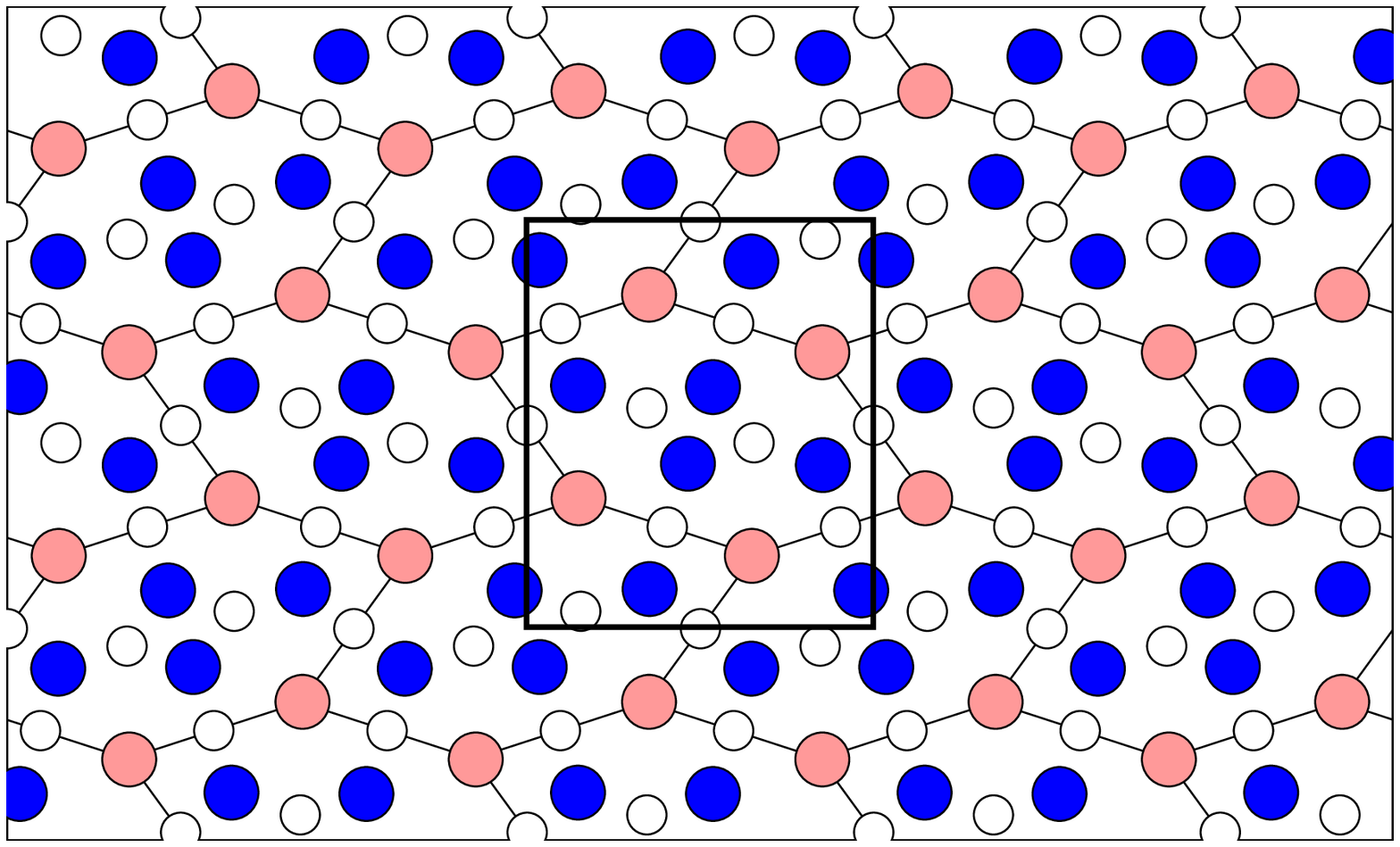}
\includegraphics[width=3in]{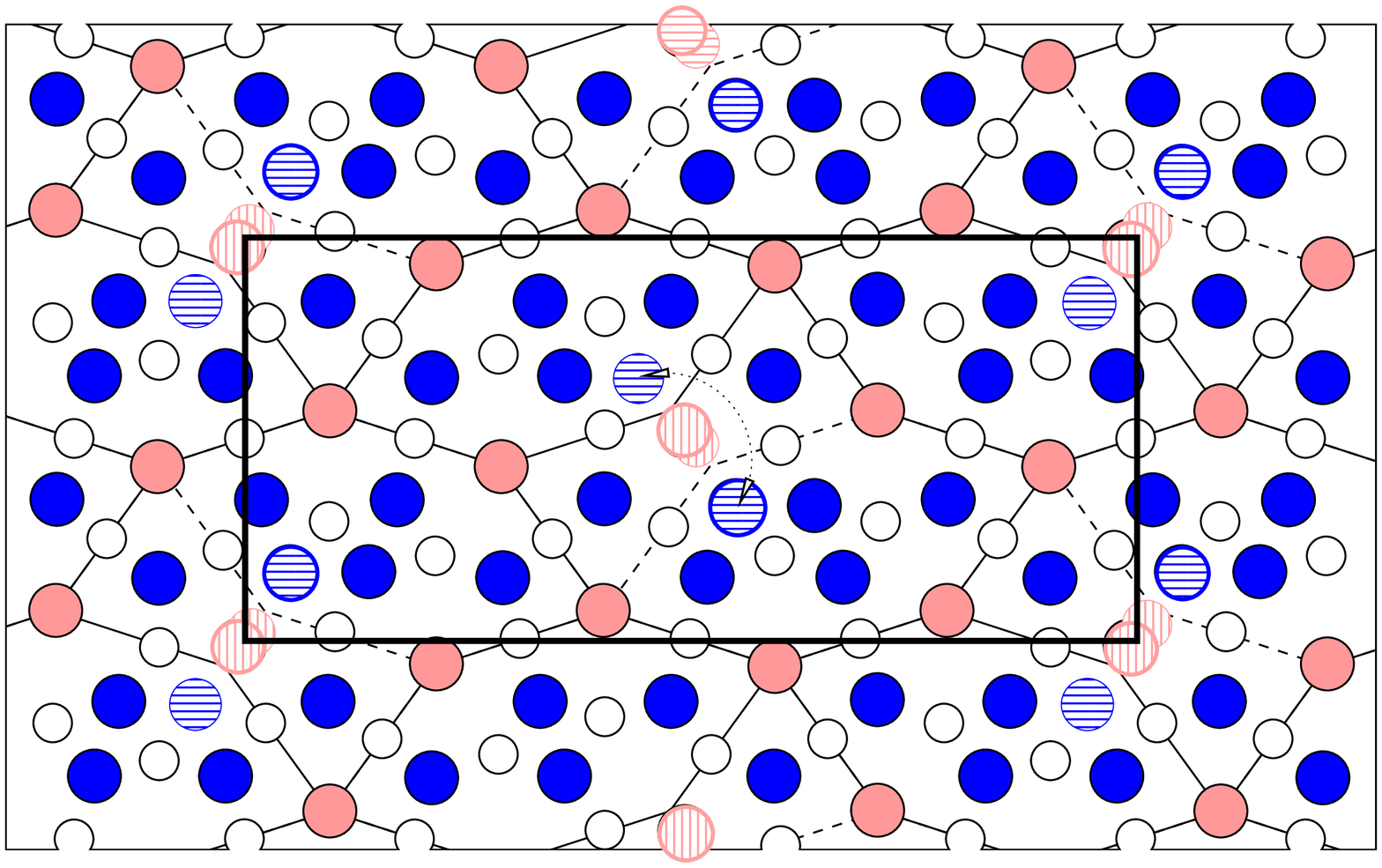}
\caption{\label{fig:expe} top: B$_4$Mg$_2$Ru$_5$ (oP22) bottom:
B$_{11}$Mg$_{5}$Ru$_{13}$ (oP62).  Color coding: Blue=B, Orange=Mg,
Green=Ru, Empty=Vacancy. Size coding: Large=upper plane, small=lower
plane. Shading indicates 1/2 atomic occupancy.  Solid lines outline
outline optimal tiling ${\cal O}$.  Dashed lines indicate possible
phason tiling flips leading to tiling ${\cal O'}$.}
\end{figure}

Two experimentally known B-Mg-Ru crystals~\cite{oP22oP62,Pearson},
 B$_4$Mg$_2$Ru$_5$ (Pearson symbol oP22, space group Pbam) and
 B$_{11}$Mg$_{5}$Ru$_{13}$ (Pearson symbol oP62, space group Pbam),
 are decagonal approximants.  These compounds form through solid state
 transformation at an annealing temperature of T$_a$=1323K.  Already a
 relationship to quasicrystals is evident in the Fibonacci numbers of
 Mg and Ru atoms in each crystal's stoichiometry.  These crystal
 structures are illustrated in Fig.~\ref{fig:expe}.  Hexagon (H) and
 boat (B) tiles~\cite{MM8} are inscribed on these figures to show how
 the structures can be interpreted as quasicrystal approximants.  The
 tile edge length $a_q$=4.5~\AA~ is known as the {\em quasilattice
 constant}.

In B$_4$Mg$_2$Ru$_5$ only the H tile appears, and it is decorated
deterministically with every site fully occupied by a unique atomic
species.  The determinism of this decoration reflects the strong size
and chemistry contrast between atomic species.  That is the sense in
which we refer to B-Mg-Ru as an {\em intrinsic} ternary.  Other
ternaries that form decagonal quasicrystals are pseudobinaries because
at least two of the elements readily substitute for each other
(e.g. Co and Ni in Al-Co-Ni).

All atoms lie on two flat layers.  The medium sized Ru atoms occupy
one layer ($z$=0), while the large and small Mg and B atoms occupy the
other ($z$=1/2).  We refer to this pair of adjacent layers as a {\em
slab}.  Slabs are stacked along the $c$ axis with a 3~\AA~ periodicity.

In our deterministic decoration, large Mg atoms occupy every tile
vertex.  Ru atoms occupy two topologically distinct sites: tile edge
midpoints ``Ru$_\alpha$'', and tile interior sites ``Ru$_\beta$''.  B
atoms form a network of pentagons and thin rhombi.  All B atoms center
trigonal prisms formed by Ru atoms~\cite{oP22oP62}. Ru$_\beta$ atoms
belong to the shared faces of the pairs of trigonal prisms inside the
tiles, while Ru$_{\alpha}$ atoms belong to shared edges of trigonal
prisms in adjacent tiles. Mg atoms serve to cap non-shared faces of
trigonal prisms. These trigonal prisms are common motif in amorphous
and related crystalline compounds~\cite{bfezry_unp,Gaskell}.

The experimentally determined atomic positions of B$_4$Mg$_2$Ru$_5$
agree very accurately with the deterministic decoration of our
idealized tiling model.  The RMS displacements of the
experimentally determined positions from the ideal positions is
0.086~\AA.  The maximum displacement is 0.14~\AA~ found for
Ru$_{\alpha}$ atoms.

\begin{figure}
\includegraphics[width=2in]{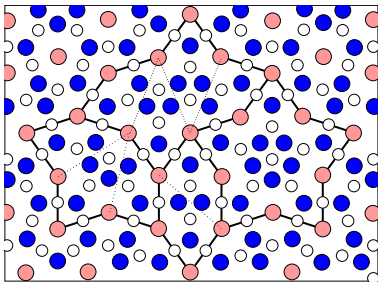}
\caption{\label{fig:model} Ideal tiles and decorations. Tiles shown
are H, B, S and E. Prototiles for the generalized 2-level
tiling~\cite{gen2L} are shown with dashed lines.}
\end{figure}


In the experimentally refined B$_{11}$Mg$_{5}$Ru$_{13}$ structure
~\cite{oP22oP62}, neither the atomic occupation nor the decomposition
into H and B tiles is uniquely determined.  The atomic occupation is
not unique because certain B and Mg atoms (shaded) have occupancy 0.5.
The decomposition into H and B tiles is not unique because an HB pair
can be interchanged by means of a {\em bowtie tile flip}~\cite{MM8}
(see dashed lines in Fig.~\ref{fig:expe}).  Tile flips are examples of
localized {\em phason} fluctuations~\cite{RTH} in quasicrystals, but
can also occur in crystalline approximants to quasicrystals where they
become discrete configurational degrees of freedom.  Because the
atomic decoration is not deterministic, there is no guide as to how to
resolve the region covered by an HB pair into its separate tiles.

Occupancy of the partially occupied sites must be highly correlated.
For example, pairs of 50\% occupied Mg sites are only 0.5~\AA~ apart.
A single Mg atom should occupy one out of the two sites at any instant
in time.  In addition, we find the nearby partially occupied B sites
are strongly correlated with the Mg position.  When the B atom
occupies its preferred site relative to the Mg atom, the HB pair
resolves uniquely into an H tile and a B tile, each decorated as shown
in Fig.~\ref{fig:model}.

Thus we believe the experiment reveals partial occupancy as a result
of tile flip disorder.  This disorder could be planar in nature, with
the HB pairs resolved randomly within a single slab that is then
stacked with perfect 3~\AA~ periodicity along the $c$ axis.
Alternatively (and more likely, as we show below) the disorder could
be truly three dimensional, with an HB pair in a given 3~\AA~ slab
possibly resolving oppositely to a BH pair in an adjacent slab,
introducing a tiling flip between slabs.

H and B tiles alone can cover the infinite plane
quasiperiodically.  However, other tile types beyond H and B are
possible (see Fig.~\ref{fig:model}).  To generate the full set of
tiles we use, start with an H tile and add a bowtie to generate a B
tile.  Adding an additional bowtie to a B tile generates either a
star (S) tile or else the tile type we call E.  This process can be
continued indefinitely creating ever larger tiles, leading to a
generalization of the set of 2-level tilings~\cite{gen2L}.  All
additional tiles in this family can be created from the set of
prototiles illustrated with dashed lines in Fig.~\ref{fig:model}.

For 10-fold symmetric quasiperiodic tilings, the composition is
rigidly fixed by the ideal atomic decorations of the tiles.  The
Boat:Hexagon ratio of $\tau:1$ (here $\tau=(\sqrt{5}+1)/2$ is the
golden mean) corresponds to fractions of atomic species as
$x_B=1/\tau^2=0.382$, $x_{Mg}=1/(\tau^2+2\tau)=0.171$ and
$x_{Ru}=\tau^2/(\tau^2+2\tau)=0.447$.  This ideal composition is about
0.1\% richer in B and Ru than the composition of the
B$_{11}$Mg$_5$Ru$_{13}$ crystal.  Owing to the the intrinsic ternary
nature of B-Mg-Ru, the composition cannot be adjusted to optimize
quasicrystal formation, in contrast to the case of the pseudobinaries,
where adjustments of the composition can move the Fermi level to a
pseudogap, for example.  If we wished to make comparable adjustments
for B-Mg-Ru it would be necessary to move to a quaternary
(pseudoternary) system such as B-(Mg,Zr)-Ru.

Given our quasicrystal model we study structural stability by
calculating cohesive energies in the B-Mg-Ru ternary system. Our
first-principles total energy calculations employ electronic density
functional theory using the plane-wave program VASP~\cite{VASP}.  We
use the GGA approximation with the VASP-supplied PAW
potentials~\cite{KJ_PAW}.  Our energy cutoff and k-point densities
achieve convergence of energy to an accuracy of better than 1
meV/atom.  For all structures examined we relax atomic positions and
lattice parameters.  We consider all known binary and ternary crystal
structures~\cite{Pearson}, a large number of hypothetical structures
drawn from chemically similar alloy systems, and 70 different quasicrystal
approximants.  Our methods are described in more detail in
Ref.~\cite{bfezry_unp} and quantitative cohesive energy data is
available on the WWW~\cite{alloy_home}.

These calculations exactly reproduce the known binary phase diagrams
of B-Mg and Mg-Ru in the sense that all known stable phases lie on the
convex hull of energy versus compositions, and all hypothetical
structures lie above the convex hull.  For B-Ru, all unknown structures
correctly lie above, and most known structures lie on the convex hull.
However, two phases, B$_3$Ru$_7$ (presumed stable) and B$_8$Ru$_{11}$
(presumed metastable) lie far above the convex hull (by 66 and 143
meV/atom, respectively).  Introducing vacancies lowers the
energy considerably but we have not yet found structures whose energy
reaches the convex hull.  These phases require further theoretical and
experimental study.  The ternary diagram is reproduced exactly.

We resolved the partial occupancy of B$_{11}$Mg$_{5}$Ru$_{13}$ in many
different ways to find the lowest energy.  It turns out that
introduction of tile flips between adjacent slabs is favored, lowering
the total energy by 8.4 meV/atom.  The lowest energy structure is
obtained by arranging two boat and two hexagon tiles in a single slab
as shown in Fig.~\ref{fig:expe} (we call this optimal tiling ${\cal
O}$), then stacking a second slab above that differs by the tile flips
outlined by dashed lines.  The resulting tiling, which we denote
${\cal O'}$, is equivalent to the starting tiling, but reflected and
translated.  The optimal structure thus exhibits a 6~\AA~ periodicity
perpendicular to the tiling plane, an alternating sequence of ${\cal
O}$ and ${\cal O'}$ with space group Pnma.  We assign this structure
Pearson symbol oP116.

Other tilings exist within the same lattice parameters.  Indeed, a
different arrangement of two boat and two hexagon tiles in a single
slab, alternating with a partner to yield 6~\AA~ periodicity and space
group Pnma as before, has energy just 0.3 meV/atom above the ${\cal
OO'}$ structure.  We denote this structure as ${\cal QQ'}$, where
${\cal Q}$ differs from ${\cal O}$ by just one bowtie flip.

Starting from ideally decorated tiles we find significant atomic
relaxation in the tiling flip regions.  In particular, the Mg atom
displaces by over 1~\AA~ towards the bowtie center (see
Fig.~\ref{fig:expe}, bottom).  Adjacent Ru$_{\alpha}$ atoms relax towards the
now-vacant vertex.  The Ru layers become slightly non-flat.  Our
relaxed positions are all within 0.05~\AA~ of the experimentally
determined atomic positions for B$_{11}$Mg$_{5}$Ru$_{13}$.

Our calculated cohesive energies show that the known ternary
structures are highly stable (enthalpies of formation are around 350
meV/atom).  However, a great many hypothetical approximant structures
lie quite close to the convex hull, starting about 2.5 meV/atom above.
In other words, we find a cluster of many distinct but very nearly
degenerate structures in the vicinity of the ideal quasicrystal
composition.

To understand why experiments find a disordered 3~\AA-periodic
structure for B$_{11}$Mg$_{5}$Ru$_{13}$, while our calculations find
an ordered 6~\AA-periodic minimum energy state, we carry out transfer
matrix calculations.  These model the temperature dependence of the
thermodynamic ensemble of structures that differ by tiling flips.
First we construct all 2-D tilings $\alpha$ that fit within a given
periodic boundary condition in the $xy$-plane.  Each tiling describes
a slab of B-Mg-Ru with height 3~\AA~ along the $z$ axis, decorated as
in Fig.~\ref{fig:model}.  Then we consider combinations
$\alpha\beta$ in which slab $\beta$ is placed above slab $\alpha$
along the $z$ axis, for a total height of 6~\AA.  We only consider
combinations which differ by disjoint localized tile flips, so the
tilings match on all vertices neighboring the flipped vertex.
Specifically, these are the four vertices forming the outside corners
of the ``bowtie'' that flips.  For each combination $\alpha\beta$,
we calculate the relaxed total energy $U_{\alpha\beta}$.

We may think of $U_{\alpha\beta}$ as comprising: the energy of slab
$\alpha$; the energy of slab $\beta$; twice the interaction energy of
slab $\alpha$ with $\beta$. This factor of 2 arises from the periodic
boundary condition along the $z$ axis.  Within this energy model, the
total energy for a stack of $n$ layers
($\alpha_1\alpha_2\alpha_3\dots,\alpha_n$)
with periodic boundary conditions of height $n\times 3$~\AA, becomes
$E_{\alpha_1\alpha_2}+E_{\alpha_2\alpha_3}+\dots+E_{\alpha_n\alpha_1}$, 
where $E_{\alpha\beta}={{1}\over{2}}U_{\alpha\beta}$.  This approximation
neglects interactions of second- and further-neighbor slabs.

Defining the transfer matrix elements
$R_{\alpha\beta}=\exp{(-E_{\alpha\beta}/k_B T)}$, 
the partition function for a stack of $n$ layers is $Z={\rm tr}~R^n$.
In the limit of many layers, the free energy per layer
$f=-k_BT\log{\rho}$, with $\rho$ the largest eigenvalue of $R$.  Other
thermodynamic quantities such as internal energy $U$, entropy $S$ and
heat capacity $C$ are given by temperature derivatives of $f$.

\begin{figure}
\includegraphics[width=3in]{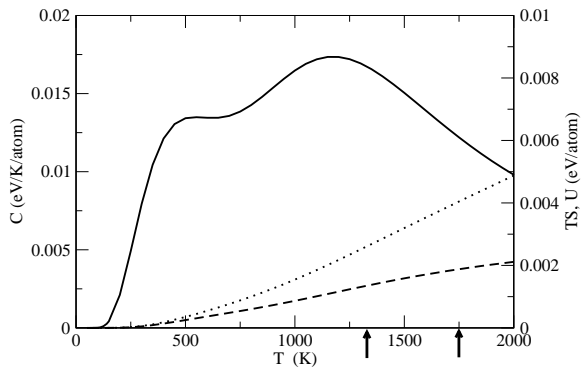}
\caption{\label{fig:Cv} Thermodynamic properties of
B$_{11}$Mg$_5$Ru$_{13}$ calculated from transfer matrix for an
infinite stack of single unit cells.  Left axis, heat capacity
$C$ (solid). Right axis, $T*S$ (dotted), $U$ (dashed).  Arrows mark
annealing temperature T$_a$ and estimated melting temperature T$_m$.}
\end{figure}

The broad peak in Fig.~\ref{fig:Cv} illustrates the heat capacity per
atom for the ensemble of structures that match the lattice parameters
of B$_{11}$Mg$_5$Ru$_{13}$ in the $xy$ plane.  At low temperature the
structure is locked into the minimum energy configuration ${\cal
OO'}$.  It has a vertical periodicity of 6~\AA~ because tiles flip
back and forth with perfect regularity.  A small peak around T=500K
indicates the onset of the alternate ${\cal QQ'}$ structure.  The heat
capacity peaks around the annealing temperature T$_a$ and well below
our guess at a likely melting temperature T$_m\approx$1750K.  We also
plot the entropy multiplied by temperature, a measure of the free
energy reduction due to tiling fluctuations.  At the annealing
temperature $TS$ is around 2.6 meV/atom.  At the same temperature,
these fluctuations increase the internal energy $U$ by around 1.3
meV/atom.

From the eigenvector of the transfer matrix we determine that about
69\% of the slabs $\alpha$ occurring in equilibrium are of the optimal
type $\alpha = {\cal O}$ or ${\cal O'}$.  Starting from a slab of
type ${\cal O}$, the next slab above is of type ${\cal O'}$ with
probability around 75\%.  This leads to a persistence length for the
6~\AA~ periodic ${\cal OO'}$ sequence of 12 slabs, or 36~\AA.  As a
result, there will be no Bragg peak associated with 6~\AA~
periodicity, but there should be pronounced diffuse scattering.

Does the broad peak in Fig.~\ref{fig:Cv} indicate a genuine phase
transition?  With finite extent in the $xy$ plane, our transfer matrix
describes an effectively one-dimensional system that is incapable of a
genuine phase transition.  A phase transition might exist in the
thermodynamic limit of infinite extent in the $xy$ plane.  This cannot
be done using ab-initio methods, which are already strained by the
demands of the 116 atom double slabs of the ${\cal OO'}$ model.
Instead, we approximated the energetics using a crude ``tile
Hamiltonian''~\cite{Tham,AW} that enforces tiling constraints and
assigns an energy preference for tiling flips between adjacent slabs.
We evaluate the tile Hamiltonian energies for supercells of the basic
B$_{11}$Mg$_5$Ru$_{13}$ structure and find that the heat capacity peak
narrows and diverges as system size grows.  The existence of a genuine
phase transition seems clear, although we do not know its temperature
very accurately owing to our crude tile Hamiltonian approximation.

Could tiling disorder, whose prevalence at high temperature we have
demonstrated, stabilize the quasicrystalline phase
entropically~\cite{entropic}?  To answer we must compare the entropy
difference between the quasicrystal and competing crystal phases with
corresponding energy differences. Substituting larger approximants for
the true quasicrystal, we find the quasicrystal is high in energy by
2.5 meV/atom.  The random tiling hypothesis~\cite{RTH} suggests the
entropy of the quasicrystal is greater than that of small
approximants.  We already found $TS$ of 2.6 meV/atom for the
B$_{11}$Mg$_5$Ru$_{13}$ phase, but we need the entropy {\em
difference} between that and the quasicrystal.  Since we cannot
directly calculate the quasicrystal entropy, we take 2.6 meV/atom as
indicating the magnitude of the probable difference in $TS$ at
temperature $T_a$.  This rises to 4.0 meV/atom at $T_m$.

Since both energy and entropy are of comparable magnitude, we have
verified from first-principles the {\em feasibility} of entropic
stabilization.  We predict that quasicrystalline B-Mg-Ru is on the
verge of stability, so effects left out of our considerations could
tip the balance.  A better estimate of the quasicrystal entropy could
be achieved by finding an accurate tile Hamiltonian then evaluating
the entropy of this model for large approximants.  We should
include the increase in internal energy $U$ caused by tile fluctuations.
Further, we should include atomic vibrations (phonons) which
contribute their own entropy and also can be expected to modify the
values of terms in the tile Hamiltonian.

In the event that further theoretical (or experimental) study shows
the B-Mg-Ru decagonal phase is not stable at high temperature, moving
to a quaternary system (e.g. substituting a small fraction of Mg with
Sc or Zr) could stabilize it by raising the energy of competing
crystal phases relative to the quasicrystal.  Regardless of the
stability at high temperature, we expect the ternary quasicrystal to
occur metastably at low temperature, because the energy differences
between crystalline and quasicrystalline structures are small, and the
high entropy of the quasicrystal implies there are a large number of
structures into which the system could freeze out of equilibrium.
Thus, we predict the occurrence of decagonal quasicrystals in
B$_{38}$Mg$_{17}$Ru$_{45}$.

We acknowledge useful discussions with C.L. Henley, Marian
Kraj\v{c}i and Yang Wang. This work was supported in part by NSF grant
DMR-0111198.

\bibliography{bmgru}

\begin{thebibliography}{19}
\expandafter\ifx\csname natexlab\endcsname\relax\def\natexlab#1{#1}\fi
\expandafter\ifx\csname bibnamefont\endcsname\relax
  \def\bibnamefont#1{#1}\fi
\expandafter\ifx\csname bibfnamefont\endcsname\relax
  \def\bibfnamefont#1{#1}\fi
\expandafter\ifx\csname citenamefont\endcsname\relax
  \def\citenamefont#1{#1}\fi
\expandafter\ifx\csname url\endcsname\relax
  \def\url#1{\texttt{#1}}\fi
\expandafter\ifx\csname urlprefix\endcsname\relax\def\urlprefix{URL }\fi
\providecommand{\bibinfo}[2]{#2}
\providecommand{\eprint}[2][]{\url{#2}}

\bibitem[{\citenamefont{Mihalkovi\v{c} and
  Widom}(2004{\natexlab{a}})}]{bfezry_unp}
\bibinfo{author}{\bibfnamefont{M.}~\bibnamefont{Mihalkovi\v{c}}}
  \bibnamefont{and} \bibinfo{author}{\bibfnamefont{M.}~\bibnamefont{Widom}},
  \bibinfo{journal}{sub. to Phys. Rev. B}
  (\bibinfo{year}{2004}{\natexlab{a}}), \bibinfo{note}{cond-mat/0405298}.

\bibitem[{\citenamefont{Mihalkovi\v{c} and Widom}(2004{\natexlab{b}})}]{mrs03}
\bibinfo{author}{\bibfnamefont{M.}~\bibnamefont{Mihalkovi\v{c}}}
  \bibnamefont{and} \bibinfo{author}{\bibfnamefont{M.}~\bibnamefont{Widom}}, in
  \emph{\bibinfo{booktitle}{Proceedings of the MRS}}, edited by
  \bibinfo{editor}{\bibfnamefont{E.}~\bibnamefont{Belin-Ferre}}
  (\bibinfo{year}{2004}{\natexlab{b}}), vol. \bibinfo{volume}{805}.

\bibitem[{\citenamefont{Zhu and Henley}(2000)}]{Henley_B}
\bibinfo{author}{\bibfnamefont{W.-J.} \bibnamefont{Zhu}} \bibnamefont{and}
  \bibinfo{author}{\bibfnamefont{C.~L.} \bibnamefont{Henley}},
  \bibinfo{journal}{Europhys. Lett.} \textbf{\bibinfo{volume}{51}},
  \bibinfo{pages}{133} (\bibinfo{year}{2000}).

\bibitem[{\citenamefont{Kimura}(1993)}]{Kimura}
\bibinfo{author}{\bibfnamefont{R.}~\bibnamefont{Kimura}},
  \bibinfo{journal}{Mat. Sci. Eng. B} \textbf{\bibinfo{volume}{19}},
  \bibinfo{pages}{67} (\bibinfo{year}{1993}).

\bibitem[{\citenamefont{Boustani et~al.}(1996)\citenamefont{Boustani, Quandt,
  and Kramer}}]{Boustani}
\bibinfo{author}{\bibfnamefont{I.}~\bibnamefont{Boustani}},
  \bibinfo{author}{\bibfnamefont{A.}~\bibnamefont{Quandt}}, \bibnamefont{and}
  \bibinfo{author}{\bibfnamefont{P.}~\bibnamefont{Kramer}},
  \bibinfo{journal}{Europhys. Lett.} \textbf{\bibinfo{volume}{36}},
  \bibinfo{pages}{583} (\bibinfo{year}{1996}).

\bibitem[{\citenamefont{Weygand and Verger-Gaugry}(1995)}]{Weygand}
\bibinfo{author}{\bibfnamefont{D.}~\bibnamefont{Weygand}} \bibnamefont{and}
  \bibinfo{author}{\bibfnamefont{J.-L.} \bibnamefont{Verger-Gaugry}},
  \bibinfo{journal}{C. R. Acad. Sci. II} \textbf{\bibinfo{volume}{320}},
  \bibinfo{pages}{253} (\bibinfo{year}{1995}).

\bibitem[{\citenamefont{Steurer et~al.}(1994)\citenamefont{Steurer, Haibach,
  Zhang et~al.}}]{other_deca}
\bibinfo{author}{\bibfnamefont{W.}~\bibnamefont{Steurer}},
  \bibinfo{author}{\bibfnamefont{T.}~\bibnamefont{Haibach}},
  \bibinfo{author}{\bibfnamefont{B.}~\bibnamefont{Zhang}},
  \bibnamefont{et~al.}, \bibinfo{journal}{J. Phys.: Cond. Mat.}
  \textbf{\bibinfo{volume}{6}}, \bibinfo{pages}{613} (\bibinfo{year}{1994}).

\bibitem[{\citenamefont{Mihalkovi\v{c} et~al.}(2002)}]{MM8}
\bibinfo{author}{\bibfnamefont{M.}~\bibnamefont{Mihalkovi\v{c}}}
  \bibnamefont{et~al.}, \bibinfo{journal}{Phys. Rev. B}
  \textbf{\bibinfo{volume}{65}}, \bibinfo{pages}{104205}
  (\bibinfo{year}{2002}).

\bibitem[{\citenamefont{Widom et~al.}(1987)\citenamefont{Widom, Strandburg, and
  Swendsen}}]{entropic}
\bibinfo{author}{\bibfnamefont{M.}~\bibnamefont{Widom}},
  \bibinfo{author}{\bibfnamefont{K.~J.} \bibnamefont{Strandburg}},
  \bibnamefont{and} \bibinfo{author}{\bibfnamefont{R.~H.}
  \bibnamefont{Swendsen}}, \bibinfo{journal}{Phys. Rev. Lett.}
  \textbf{\bibinfo{volume}{58}}, \bibinfo{pages}{706} (\bibinfo{year}{1987}).

\bibitem[{\citenamefont{Schweitzer and Jung}(1985)}]{oP22oP62}
\bibinfo{author}{\bibfnamefont{K.}~\bibnamefont{Schweitzer}} \bibnamefont{and}
  \bibinfo{author}{\bibfnamefont{W.}~\bibnamefont{Jung}}, \bibinfo{journal}{Z.
  Anorg. Allg. Chemie} \textbf{\bibinfo{volume}{530}}, \bibinfo{pages}{127}
  (\bibinfo{year}{1985}).

\bibitem[{\citenamefont{Villars}(1997)}]{Pearson}
\bibinfo{author}{\bibfnamefont{P.}~\bibnamefont{Villars}},
  \emph{\bibinfo{title}{Pearson's Handbook, Desk Edition}}
  (\bibinfo{publisher}{ASM International}, \bibinfo{address}{Materials Park,
  Ohio}, \bibinfo{year}{1997}).

\bibitem[{\citenamefont{Gaskell}(1978)}]{Gaskell}
\bibinfo{author}{\bibfnamefont{P.}~\bibnamefont{Gaskell}},
  \bibinfo{journal}{Nature} \textbf{\bibinfo{volume}{276}},
  \bibinfo{pages}{484} (\bibinfo{year}{1978}).

\bibitem[{\citenamefont{Henley}(1998)}]{gen2L}
\bibinfo{author}{\bibfnamefont{C.~L.} \bibnamefont{Henley}}, in
  \emph{\bibinfo{booktitle}{Quasicrystals}}, edited by
  \bibinfo{editor}{\bibfnamefont{S.}~\bibnamefont{Takeuchi}} \bibnamefont{and}
  \bibinfo{editor}{\bibfnamefont{T.}~\bibnamefont{Fujiwara}}
  (\bibinfo{publisher}{World Scientific, Singapore}, \bibinfo{year}{1998}),
  p.~\bibinfo{pages}{27}.

\bibitem[{\citenamefont{Henley}(1991)}]{RTH}
\bibinfo{author}{\bibfnamefont{C.~L.} \bibnamefont{Henley}}, in
  \emph{\bibinfo{booktitle}{Quasicrystals: the state of the art}}, edited by
  \bibinfo{editor}{\bibfnamefont{D.~P.} \bibnamefont{DiVincenzo}}
  \bibnamefont{and} \bibinfo{editor}{\bibfnamefont{P.~J.}
  \bibnamefont{Steinhardt}} (\bibinfo{publisher}{World Scientific},
  \bibinfo{address}{Singapore}, \bibinfo{year}{1991}), pp.
  \bibinfo{pages}{429--524}.

\bibitem[{\citenamefont{Kresse and Hafner}(1993)}]{VASP}
\bibinfo{author}{\bibfnamefont{G.}~\bibnamefont{Kresse}} \bibnamefont{and}
  \bibinfo{author}{\bibfnamefont{J.}~\bibnamefont{Hafner}},
  \bibinfo{journal}{Phys.\ Rev. B} \textbf{\bibinfo{volume}{47}},
  \bibinfo{pages}{RC558} (\bibinfo{year}{1993}).

\bibitem[{\citenamefont{Kresse and Joubert}(1999)}]{KJ_PAW}
\bibinfo{author}{\bibfnamefont{G.}~\bibnamefont{Kresse}} \bibnamefont{and}
  \bibinfo{author}{\bibfnamefont{D.}~\bibnamefont{Joubert}},
  \bibinfo{journal}{Phys. Rev. B} \textbf{\bibinfo{volume}{59}},
  \bibinfo{pages}{1758} (\bibinfo{year}{1999}).

\bibitem[{\citenamefont{Mihalkovi\v{c} and
  Widom}(2004{\natexlab{c}})}]{alloy_home}
\bibinfo{author}{\bibfnamefont{M.}~\bibnamefont{Mihalkovi\v{c}}}
  \bibnamefont{and} \bibinfo{author}{\bibfnamefont{M.}~\bibnamefont{Widom}},
  \bibinfo{howpublished}{web site} (\bibinfo{year}{2004}{\natexlab{c}}),
  \bibinfo{note}{http://alloy.phys.cmu.edu}.

\bibitem[{\citenamefont{Mihalkovi\v{c}
  et~al.}(1996)\citenamefont{Mihalkovi\v{c}, Zhu, Henley et~al.}}]{Tham}
\bibinfo{author}{\bibfnamefont{M.}~\bibnamefont{Mihalkovi\v{c}}},
  \bibinfo{author}{\bibfnamefont{W.-J.} \bibnamefont{Zhu}},
  \bibinfo{author}{\bibfnamefont{C.~L.} \bibnamefont{Henley}},
  \bibnamefont{et~al.}, \bibinfo{journal}{Phys. Rev. B}
  \textbf{\bibinfo{volume}{53}}, \bibinfo{pages}{9021} (\bibinfo{year}{1996}).

\bibitem[{\citenamefont{Al-Lehyani and Widom}(2003)}]{AW}
\bibinfo{author}{\bibfnamefont{I.}~\bibnamefont{Al-Lehyani}} \bibnamefont{and}
  \bibinfo{author}{\bibfnamefont{M.}~\bibnamefont{Widom}},
  \bibinfo{journal}{Phys. Rev. B} \textbf{\bibinfo{volume}{67}},
  \bibinfo{pages}{014204} (\bibinfo{year}{2003}).

\end{thebibliography}

\end{document}